\documentclass{ametsocV6.1}

\usepackage{fancyhdr}

\title{Assimilation of SMAP Observations Over Land Improves the Simulation and Prediction of Tropical Cyclone Idai\\ {\it This manuscript has been submitted to Monthly Weather Review. Copyright in this manuscript may be transferred without further notice.}}

\authors{Jana Kolassa,\aff{a,b}\correspondingauthor{Jana Kolassa, jana.kolassa@nasa.gov} 
Manisha Ganeshan,\aff{c,d} 
Erica L. McGrath-Spangler,\aff{a,c} 
Oreste Reale,\aff{a,b} 
Rolf H. Reichle,\aff{a} 
Sara Q. Zhang, \aff{a,e}
}

\affiliation{\aff{a}{Global Modeling and Assimilation Office,
NASA Goddard Space Flight Center
Greenbelt, MD, USA}\\
\aff{b}{Science Systems and Applications, Inc., Lanham, MD, USA}\\
\aff{c}{Morgan State University, Baltimore, MD, USA}\\
\aff{d}{Climate and Radiation Laboratory,
NASA Goddard Space Flight Center
Greenbelt, MD, USA}
\aff{e}{Science Applications International Corporation, Reston, VA, USA}
}
\abstract{Soil moisture conditions can influence the evolution of a tropical cyclone (TC) that is partially or completely over land. Hence, better constraining soil moisture initial conditions in a numerical weather prediction model can potentially improve predictions of TC evolution near or over land. This study examines the impact of assimilating observations from the NASA Soil Moisture Active Passive (SMAP) mission into the NASA Goddard Earth Observing System (GEOS) global weather model on the prediction of South-West Indian Ocean TC Idai (2019). Two sets of retrospective forecasts of TC Idai are compared in an Observing System Experiment framework: (i) forecasts initialized from an analysis that is comparable to the GEOS operational analysis and (ii) forecasts initialized from an analysis that additionally assimilates SMAP brightness temperature observations over land. Results indicate that SMAP assimilation leads to pronounced improvements in the representation of TC Idai structure and prediction of its intensity and track. The wind speed radius (a measure for TC compactness) is reduced by up to 18\% in the analysis with SMAP assimilation relative to the control experiment without SMAP assimilation. The forecast intensity error, measured against the observed intensity, is reduced by up to 23\%.  The forecast along-track error is reduced by up to 34\%, indicating a more accurate propagation speed, while the impact of SMAP assimilation on the forecast cross-track error is neutral. These results provide a valuable demonstration that SMAP assimilation can have a highly beneficial impact on TC prediction in global weather forecast models.}

\begin{document}

\maketitle

\pagestyle{fancy}
\fancyhf{}
\lhead{{\it This manuscript has been submitted to Monthly Weather Review. Copyright in this manuscript may be transferred without further notice.}}

%
%
%
\statement

This manuscript summarizes the results from a study demonstrating that incorporating observations from the Soil Moisture Active Passive (SMAP) mission into a global weather forecast model is able to meaningfully improve the prediction of tropical cyclone Idai intensity and track. The work highlights the value SMAP observations in improving predictions of extreme weather events and presents a significant step towards being better able to prepare for and mitigate the socio-economic impact of extreme events.

%
%
%

%


 \section{Introduction}
 \label{sec:introduction}

 Over warm ocean waters in the deep tropics, the  maintenance of a mature tropical cyclone (TC) stems from a highly unstable balance between the low-level convergent flow advecting enthalpy flux into the system and the upward expansion, heat release and upper-level divergent flow that counteracts the process (e.g., \citet{Emanuel2003}). From a thermodynamic and operational forecast perspective, it is generally assumed that the primary forcings that alter the TC energy balance are 1) the thermodynamic disequilibrium between the ocean's surface temperature and the near-surface atmospheric temperature, and 2) the atmospheric horizontal temperature gradients and the associated vertical wind shear. Within the same approximating set of assumptions, the land's prominent role in TC wind decay is the mechanical action of increased surface roughness \citep{Kaplan2001} and the asymmetries induced by the roughness contrast between land and ocean \citep{Wong2007}. An increase in low-level convergence that is not counteracted by increased upper-level divergence leads to rising center pressure and eventual cyclone dissipation. 
 
 In the past, little attention was paid to the land's thermodynamic state, under the assumption that land is intrinsically too dry to sustain a TC and that moisture contributions from land, if any, are irrelevant. In an operational context, this thinking remains the working hypothesis. However, some storms have been noted to intensify over sandy desert areas, particularly over northern Australia. A seminal work by \cite{Emanuel2008} analyzed TC Abigail (2001) with the aid of a relatively simple land-surface model and suggested that the peculiar nature of the sandy soil in the region, its remarkable thermal capacity and conductivity, together with some wetting that occurred at the onset of the storm's passage, might have produced fluxes strong enough to enable re-intensification. Over the following years, many more storms were observed to undergo surprisingly long phases over land with modest or no weakening. For example, \cite{Arndt2009, Evans2011} and \citet{Kellner2012} investigated TC Erin (2007), finding evidence consistent with \cite{Emanuel2008} and further corroborating the idea that soil moisture can indeed delay TC dissipation or even produce some temporary strengthening. 
 
More recently, increased attention has been placed on the thermodynamic role of land, via soil moisture, in contributing to the persistence of TCs after landfall (e.g., \cite{Andersen2013, Andersen2014}). One particularly interesting TC that flared intense debate about the relevance of land conditions was Tropical Storm Bill (2015), which made landfall in Texas at 1645 UTC 16 June 2015 as a 50-kt tropical storm following weeks of extended rainfall in the region.
Despite never reaching hurricane strength, and after being downgraded to a tropical depression at 0600 UTC 17 June 2015, Bill survived over land with a clearly identifiable tropical circulation for more than 48 hours after landfall, being declared extratropical only at 1800 UTC 18 June 2015 \citep{nhc2015}. \citet{Andersen2017} argued that elevated soil moisture played a crucial role in Bill's maintenance, noting the persistence of a TC-type circulation over land, and the similarity with the aforementioned TC Erin (2007). Furthermore, \cite{Wakefield2021} and \cite{Brauer2021} confirmed the maintenance of Bill's tropical characteristics, even in the precipitation structure, hundreds of kilometers inland. The storm's exceptional resilience to land drag, particularly remarkable because of the storm's relatively weak winds, further emphasizes the prominent role played by elevated soil moisture caused by preceding, extended rainfall from unrelated weather systems. 
 
Despite several studies having demonstrated a prominent role of soil moisture in the maintenance of a variety of tropical systems over land \citep[e.g.][]{Nair2019,Zhang2019,Yoo2020,Shepherd2021,Zhang2021,Zhu2022,Osuri2020}, soil moisture information is under-utilized in several global operational weather forecast systems. Historically, progress was hindered by a sparsity of high-quality, global soil moisture observations as well as outdated land-surface parameterizations that were designed to serve as boundary conditions for the atmosphere and to compensate for atmospheric errors. With significant advances in land surface modelling and the emergence of high-quality land surface observations, land data assimilation in numerical weather prediction (NWP) models has become more feasible. Initial studies examined the assimilation of screen-level observations to indirectly constrain land surface states \citep{Mahfouf2011}, but with the emergence of high-quality satellite soil moisture observations, some NWP centers have moved to directly constraining soil moisture states. In particular, space-borne L-band (1.4 GHz) brightness temperature observations from the Soil Moisture Ocean Salinity (SMOS; \cite{kerr2010}) and Soil Moisture Active Passive (SMAP; \cite{entekhabi2010}) missions -- both missions dedicated to observing surface soil moisture -- offer a unique opportunity to improve the land surface states in NWP models. For example, soil moisture retrievals from SMOS and the Advanced Scatterometer (ASCAT) are routinely assimilated in the European Center for Medium-Range Weather Forecasts' (ECMWF) operational analysis \citep{derosnay2013,rodriguez2019,munoz2019}, and SMAP soil moisture retrievals are operationally assimilated in the US Air Force forecast system \citep{wegiel2020}. The impact of assimilating SMAP L-band brightness temperatures in the Canadian Land Data Assimilation System \citep{Carrera2015} has been investigated by \cite{Carrera2019}, who found reduced biases in near-surface states and locally improved precipitation forecasts. A recent study by \cite{reichle2023} has demonstrated the positive impact of assimilating SMAP brightness temperatures on global forecasts of near-surface air temperature and humidity in the NASA Goddard Earth Observing System (GEOS). 

Land data assimilation (DA) offers the opportunity to leverage the longer memory of the land surface relative to the atmosphere for increased predictability and thus improved forecasts. While the impact of soil moisture DA in global forecast systems was found to be positive, the global average impact was generally small \citep{Carrera2019,munoz2019,reichle2023}. This is expected, because modern data assimilation systems typically ingest more than $10^7$ observations per cycle, and the marginal impact of assimilating additional observations is limited to the independent information that they can provide. Furthermore, the land influences the atmosphere only in certain locations and under certain conditions, which further mutes the impact of land DA when assessed at the global scale. A potentially more promising path forward to quantify the true value of land DA in NWP systems, and the one chosen in this study, is an event-based approach that assesses the impact of land DA for a particular weather event. 

The main and novel objective of the present work is to examine if the assimilation of SMAP observations over land in a global NWP system results in a discernible positive impact on TC forecast skill when part or all of the TC circulation is over land. While only a subset of TCs make landfall, and the impact of SMAP DA on global forecast skill may be small, the forecast improvement for a single TC in proximity to land or after landfall has important societal implications. For this reason, our study focuses on TC Idai, which ravaged Mozambique and Zimbabwe in 2019 (section \ref{sec:tc_idai}) and is an example of a catastrophic system that interacted very strongly with land. The impact of the SMAP assimilation on the representation of the storm's evolution, track, structure, and intensity is assessed with an observing system experiment (OSE) framework that uses the NASA GEOS data assimilation and forecast system. The paper is organized as follows.  Section \ref{sec:tc_idai} provides a synoptic history of Idai and  illustrates its uniqueness from the point of view of its unusual track and the prominent role played by land. Section \ref{sec:data_and_methodology} presents the methodology, defines the specifics of the OSE framework, the data, as well as the details of the SMAP assimilation, the experiment setup and validation. Section \ref{sec:results_tc} presents the impact of SMAP DA on forecasts of TC Idai, and Section \ref{sec:results_surface_mechanisms} examines the underlying mechanisms leading to forecast improvements. Section \ref{sec:conclusions} summarizes the conclusions of this work.

 \section{Tropical Cyclone Idai: synoptic history and uniqueness}
 \label{sec:tc_idai}

\begin{figure}[t]
\centering
\noindent\includegraphics[width=0.9\textwidth,angle=0]{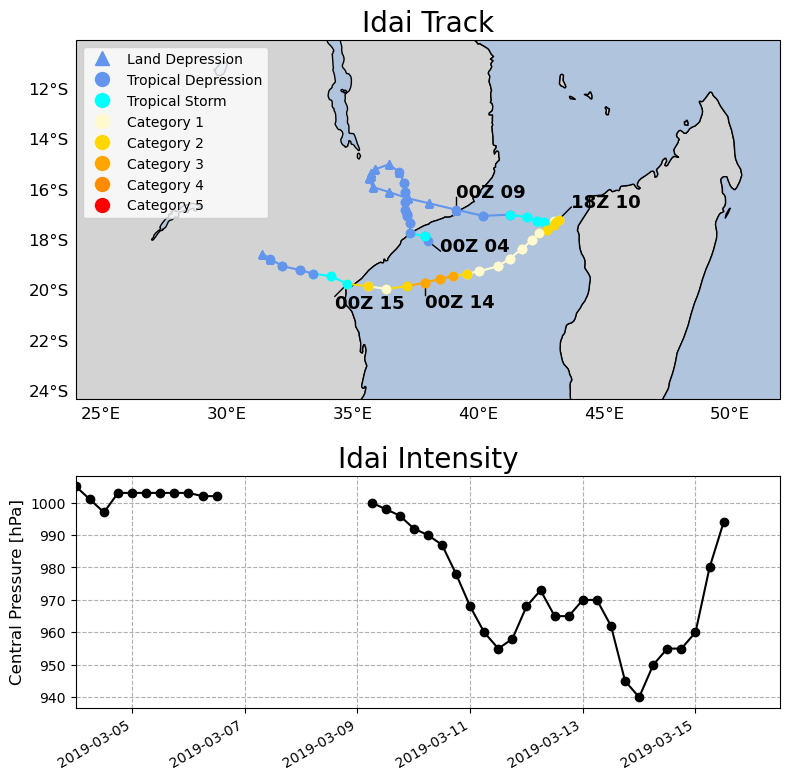}\\
\caption{(top) Track of TC Idai between 0000 UTC 4 March 2019 until its dissipation over land at 1200 UTC 16 March 2019, as observed by the International Best Track Archive for Climate Stewardship (IBTrACS, see section \ref{sec:data_and_methodology}\ref{ssec:validation}) and (bottom) central pressure. Colors in the track plot indicate the storm intensity. Dots are showing TC center location and pressure every 6 hours.  Some center pressure values between 7 and 9 March are missing because of difficulties in determining a closed circulation during Idai's vast loop over land.}
\label{TrackSLP}
\end{figure}

\begin{figure}[t]
\centering
\noindent\includegraphics[width=0.7\textwidth,angle=0]{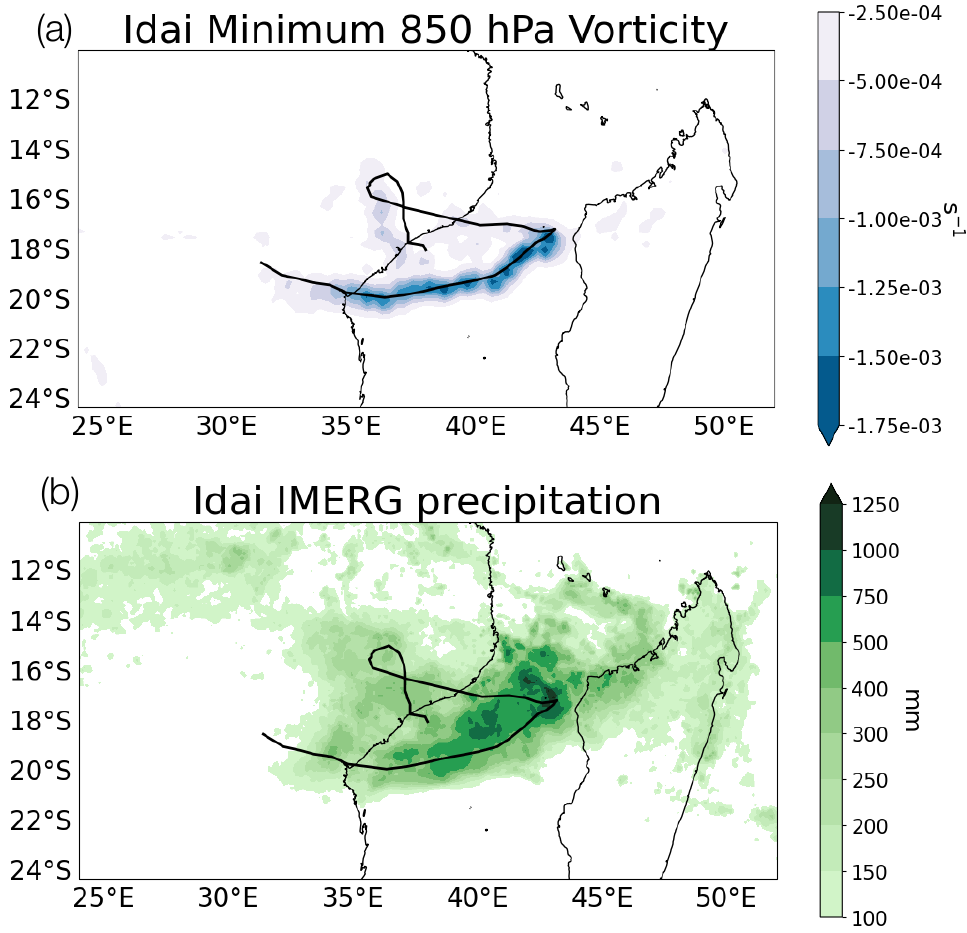}\\
\caption{(a) Minimum 850 hPa vorticity in s$^{-1}$, computed from the GEOS control analysis (CNTRL\_ANA) across Idai's lifetime, over-imposed on Idai's track. (b) Total accumulated precipitation from IMERG version 6 Final, over-imposed on Idai's track.}
\label{fig:vort_precip}
\end{figure}

\begin{figure}[t]
\centering
\noindent\includegraphics[width=0.7\textwidth,angle=0]{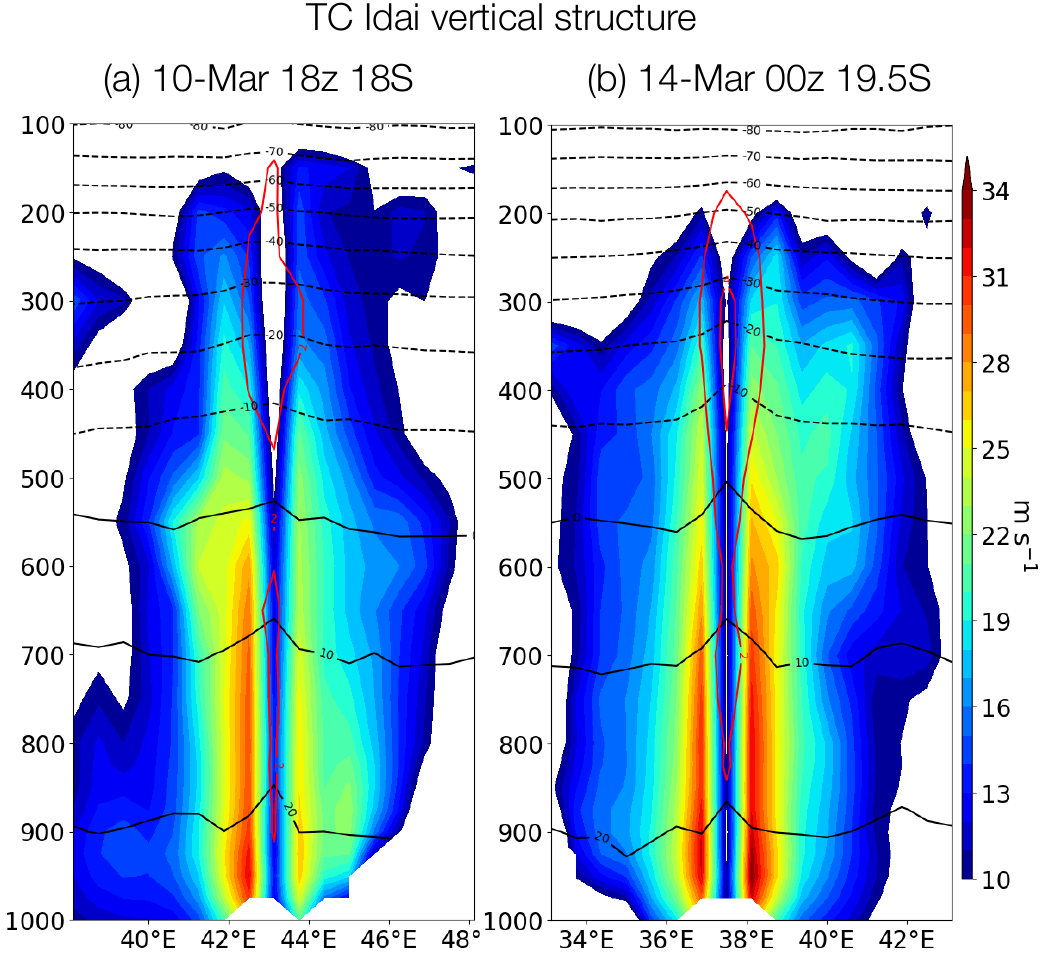}\\
\caption{ Vertical zonal cross-section of wind (shaded, m s$^{-1}$), temperature (solid black line, \textdegree C), and temperature anomaly (solid red line, \textdegree C) at (a) closest proximity to Madagascar (1800 UTC 10 March, 18 hours before the first intensity peak) and (b) at maximum intensity (0000 UTC 14 March).}
\label{fig:vstruct}
\end{figure}

The case study chosen for this article is TC Idai, which occurred in the South-West Indian Ocean in 2019 and caused a humanitarian catastrophe of immense proportions, as described, among others, by the Center for Disaster Philanthropy \citep{cdp2019}. Loss of life and damages caused by Idai  were colossal, with at least 1500 casualties and, according to the World Meteorological Organization \citep{wmo2022}, Idai was the deadliest and costliest TC in the South-West Indian Ocean basin. Casualties occurred not only in Mozambique, but also, to a lesser extent, in Madagascar, Malawi, South Africa and Zimbabwe. The World Health Organization estimated that, two months after the storm, 1.85 million people were still in need of humanitarian assistance and protection \citep{who2019}. 

From a meteorological perspective, Idai was a formidable TC with a complex and unusual life cycle that included two separate landfalls and a long persistence over land before a return to the ocean and re-intensification. Figure~\ref{TrackSLP} illustrates the track and central pressure, based on the National Oceanic and and Atmospheric Administration (NOAA) National Center for Environmental Information (NCEI) International Best Track Archive for Climate Stewardship (IBTrACS \citet{knapp2010, knapp2018}). 
Idai's precursor was spotted over the Mozambique channel on 1 March 2019, and organized into tropical depression 11 at 0600 UTC 4 March 2019, with 1003 hPa central pressure and $15~m s^{-1}$ wind speed, as reported by M\'et\'eo-France La R\'eunion (MFLR) in the first warning for this system \citep{mflr2019}.
Subsequent MFLR warnings document that the depression made landfall in Northern Mozambique around 1800 UTC and then proceeded over land, retaining a distinguishable tropical circulation and producing intense precipitation. 
However, the long loop over land was not enough to lead to a complete dissipation, even if the surface signal was very weak. In fact, the system re-emerged in the Mozambique channel on 9 March, surviving almost five days over land. Figure~\ref{fig:vort_precip}a shows the minimum 850 hPa vorticity, the relevant quantity for Southern Hemisphere TCs, computed from the GEOS control analyses (section \ref{sec:data_and_methodology}\ref{ssec:experiment_setup}). Despite the resolution limitation of the model, a weak but not negligible coherent tropical structure persisting throughout the period in which the system was over land can be seen.  Figure~\ref{fig:vort_precip}b shows total accumulated precipitation, as obtained from the Integrated Multi-satellitE Retrievals for the Global Precipitation Measurement mission (IMERG-Final version 6; \cite{tan2019}). The scale of precipitation over land during the loop is consistent with the behavior of a tropical depression remnant. 

After returning over water, the system moved eastward (Figure~\ref{TrackSLP}), steered by mid-latitude westerly flow (not shown). However, by 11 March, whilst near Madagascar, TC Idai disentangled from the westerly flow, stalled, and initiated rapid intensification with an eyewall replacement cycle, reaching a central pressure of 955 hPa (Figure~\ref{TrackSLP}) and an estimated strength comparable to a category 2 Hurricane. It then moved westward again, slightly weakening to a category 1, with central pressure rising to about 973 hPa. In the middle of the Mozambique channel, Idai then underwent a second rapid intensification cycle and reached peak intensity at 0000 UTC 14 March, with 940 hPa center pressure and winds at about 57~m s\textsuperscript{-1}. In Figure~\ref{fig:vstruct}, zonal vertical cross-sections of wind and temperature, as represented in the GEOS analyses, are displayed for 10 and 14 March. Despite the  relatively coarse (0.25 degree) model resolution, at which the storm's actual intensity is typically underestimated, a prominent vertical alignment and a warm core are evident. After reaching peak intensity, Idai's westward motion accelerated, while exhibiting small intensity fluctuations, according to the Joint Typhoon Warning Center (JTWC) report \citep{jtc2019}. 

The second,  more catastrophic landfall, occurred in Central Mozambique at 0000 UTC 15 March (Figure~\ref{TrackSLP}) with winds exceeding 45~m s\textsuperscript{-1}. 
While the storm rapidly lost intensity after landfall with winds decaying to tropical depression level as documented in the MFLR reports, the devastating effects from freshwater floods were the dominant aspects. In fact, accumulated precipitation amounts  were on the order of 100 to 300 mm for the second landfall (Figure~\ref{fig:vort_precip}b), which caused extensive flooding due to the fact that the soil was already saturated by precipitation from the first landfall. Maximum totals of accumulated precipitation for Idai's entire life cycle reached 600 mm at several locations. 

Idai is particularly relevant for this study because it originated and spent its entire lifetime (which spanned from 4 March 2019 to 16 March 2019) over a very small area, being either in the Mozambique Channel or over land
(Figure~\ref{TrackSLP}). The exceptional confinement of Idai's track, between 20\textdegree S and 14\textdegree S latitude and 30\textdegree W and 45\textdegree E longitude, and continuous constraint by  two landmasses (the eastern part of the African continent, and Madagascar) is particularly noteworthy because of Idai's relatively long lifetime. Throughout its synoptic history, Idai's circulation was affected by land. Therefore, the role played by the thermodynamic state of land, and soil moisture in particular, is expected to be highly relevant. The erratic track is also an indication of an absence of a sustained mid-tropospheric steering flow, which may have allowed for a stronger than usual role played by land-surface processes.

For the purpose of this paper, this investigation will focus on the period of TC Idai (March 4 - March 16) and on the spatial domain shown in Figure~\ref{TrackSLP} (10\textdegree S - 24\textdegree S and 24\textdegree E - 52\textdegree E), thus encompassing the domain of Idai's life cycle.
 
 \section{Data and Methodology}
 \label{sec:data_and_methodology}
 \subsection{Observing System Experiment}
 \label{ssec:ose}
 
Central to our study is the development of an Observing System Experiment framework, aimed at systematically assessing the impact of SMAP DA over land on forecasts  of TCs. Generally, an OSE framework is designed to compare a set of control model simulations that do not assimilate the observation of interest to a second set of simulations that do. The experiments are implemented using the NASA GEOS modeling and data assimilation framework and are configured to mimic the GEOS Forward Processing (FP) quasi-operational weather analysis and forecasting system.

The OSE setup consists of four experiments, two analyses and two sets of forecast simulations (Table \ref{tab:experiment_overview}). The control analysis (CNTRL\_ANA) is generated by running the GEOS atmospheric general circulation model (AGCM) with its complete atmospheric data assimilation system (ADAS), ingesting all the atmospheric observations routinely assimilated in the operational model, with the exception of TC Vital Statistics ("TC Vitals", \citet{trahan2012}). The experiment analysis (SMAP\_ANA) is identical to CNTRL\_ANA but for the additional assimilation of SMAP brightness temperature (Tb) observations over land using a weakly-coupled land analysis (section \ref{sec:data_and_methodology}\ref{ssec:LADAS}). Forecasts are produced by integrating the GEOS AGCM forward in time without any data assimilation. Each forecast is initialized from the 0000 UTC analysis and integrated forward five days (120 hours). The control forecasts (CNTRL\_FCST) are initialized from CNTRL\_ANA, while the experiment forecasts (SMAP\_FCST) are initialized from SMAP\_ANA.

To assess the impact of SMAP DA, the experiments with and without SMAP assimilation are compared in terms of the analyzed and forecast land surface and TC characteristics.

 \subsection{Data Assimilation System}
 \label{ssec:LADAS}
 
The GEOS data assimilation system used here consists of an AGCM \citep{molod2015} and an atmospheric analysis with a weakly coupled land analysis (\cite{reichle2023}; their Figure 1). The AGCM is running on a cubed-sphere grid and encompasses a range of atmospheric physics schemes as well as the Catchment land surface model \citep{koster2000}. The atmospheric data assimilation system (ADAS) uses a hybrid 4-dimensional ensemble variational (Hybrid-4dEnVar) approach, which produces an atmospheric analysis every 6 hours based on satellite and conventional observations of the atmosphere  \citep{todling2018}. The Hybrid-4DEnVar ADAS consists of a single-member, deterministic, 4-dimensional variational assimilation component and a coarser-resolution, 32-member ensemble data assimilation component for the simulation of flow-dependent background errors. 
A weakly-coupled land analysis using a land-only ensemble Kalman filter (EnKF) ingests the SMAP Tb observations to produce a 3-hourly soil moisture analysis (section~\ref{sec:data_and_methodology}\ref{ssec:SMAP-DA}). 
The soil moisture increments computed in the land analysis are applied to the AGCM component of the ADAS, while the surface meteorology simulated by the AGCM is used to force the land model. In this weakly coupled system, observation or model background error information is not shared between the atmospheric analysis and the land analysis; the atmospheric analysis produces increments only for atmospheric model states, and the land analysis produces only soil moisture and soil temperature increments. Consequently, the influence of the land surface on the atmosphere occurs primarily through the physical land-atmosphere interactions encoded in the AGCM. 

The deterministic component of the AGCM in this study is set up on the C360  cubed-sphere grid (approximately 0.25-degree resolution), permitting to a reasonable extent the simulation of a TC circulation. The coarser resolution ensemble component is run on a C90 cubed-sphere grid (1-degree resolution). Two instances of the land analysis are run, one matching the 0.25-degree AGCM resolution of the deterministic ADAS component and one matching the 1-degree AGCM resolution of the ADAS ensemble. The 0.25-degree resolution land analysis uses 24 ensemble members and applies the ensemble-average increment to the single-member (deterministic) AGCM. The 1-degree resolution land analysis uses 32 ensemble members, matching the ensemble size of the AGCM ensemble, and its land increments are applied member-wise to the AGCM ensemble. 
The weakly-coupled land analysis used here is described in detail and evaluated extensively in \cite{reichle2023}, and the reader is referred to their paper for further information. 

\subsection{SMAP Data Assimilation}
\label{ssec:SMAP-DA}
 
The SMAP mission was launched in 2015 and is equipped with an L-band (1.4 GHz) radiometer that is highly sensitive to water in the surface (0-5 cm) soil layer \citep{entekhabi2010}. SMAP collects observations of horizontal (H) and vertical (V) polarization brightness temperatures from a sun-synchronous, near-circular, polar orbit with equator crossings at 6 AM and 6 PM local time and a revisit time of 2-3 days. Here we assimilate H- and V-polarization SMAP Level 1C brightness temperatures \citep{chan2020} from both the AM and PM overpasses, which are provided as daily half-orbit files on the 36-km resolution Equal-Area Scalable Earth version 2 (EASEv2) grid \citep{brodzik2012}. 
 
The land analysis component of the data assimilation system is a variant of the SMAP Level-4 Soil Moisture (L4 SM) algorithm \citep{reichle2017} that has been configured to work on the cubed-sphere grid. An ensemble of Catchment land model simulations is used to represent model uncertainty. The ensemble spread is maintained by adding perturbations to some of the model forcing and prognostic variables \citep{reichle2017}. Once every three hours, the available SMAP Tb observations are scaled into the model's climatology and are compared with the corresponding model forecast Tbs estimated from a tau-omega radiative transfer model for L-band \citep{delannoy2013,delannoy2014}. Based on the Tb observation-minus-forecast differences, and taking into consideration the relative uncertainties of each, a spatially distributed ensemble Kalman filter (EnKF) analysis generates soil moisture increments. Here we assume an observation error of 4K for the SMAP Tbs globally. The increments are then added to the Catchment model estimates in order to make the modeled soil moisture and temperature fields more consistent with the SMAP Tb observations. 

While the SMAP Tb observations are typically only sensitive to soil moisture in the top few centimeters of the soil, the EnKF analysis also generates increments for root-zone soil moisture. This is possible by including unobserved Catchment model variables related to deeper-layer soil moisture in the EnKF state vector and calculating increments based on the simulated (ensemble) error correlations between the surface and root-zone layers \citep{reichle2017}. The EnKF state vector consists of the model prognostic variables “surface excess” (SRFEXC) and “root-zone excess” (RZEXC), with units kg m$^{-2}$, from which volumetric soil moisture in units of m$^3$ m$^{-3}$ is diagnosed for the “surface” (0-5 cm), “root-zone” (0-100 cm), and “profile” (0 cm to bedrock) layers.
 
 \subsection{Experiment Setup}
 \label{ssec:experiment_setup}
 
The OSE framework for TC Idai consists of the four model simulations discussed in section \ref{sec:data_and_methodology}\ref{ssec:ose} and detailed in Table \ref{tab:experiment_overview}. Critical for correctly assessing the impact of SMAP DA on the TC Idai predictions is to allow for a sufficient spin-up of the simulations, such that all observed differences can be attributed to the assimilation of SMAP. This holds particularly true for the simulated land surface states, which evolve more slowly than the atmospheric states. To this end, the two analyses (CNTRL\_ANA and SMAP\_ANA) are conducted from 15 December 2018 to 23 March 2019, allowing approximately two months of simulations before the beginning of the TC Idai life cycle. Additionally, the 15 December 2018 land states in CNTRL\_ANA and SMAP\_ANA are initialized from multi-year, land-only model simulation without and with SMAP DA, respectively. 

\begin{table}[t]
\caption{Overview of the four simulations conducted.}\label{tab:experiment_overview}
\begin{center}
\begin{tabular}{lp{5cm}p{4cm}p{3cm}}
\hline\hline
Experiment Name & Observations assimilated & Simulation period & Initialization\\
\hline
 CNTRL\_ANA & GEOS-FP operational atmospheric observations & 4-16 Mar 2019  & spin-up analysis (see text)\\
 SMAP\_ANA & GEOS-FP operational atmospheric observations + SMAP brightness temperatures & 4-16 Mar 2019 & spin-up analysis (see text) \\
 CNTRL\_FCST & none & 5-day forecasts initialized daily at 0000 UTC during 4-16 Mar 2019 & CNTRL\_ANA  \\
 SMAP\_FCST & none & 5-day forecasts initialized daily at 0000 UTC during 4-16 Mar 2019 & SMAP\_ANA \\
 \hline
\end{tabular}
\end{center}
\end{table}
 
 The Hybrid-4DEnVar data assimilation system used in this study is computationally expensive. In order to mitigate the computational cost of the spin-up, the spin-up period was split into two phases. Spinup during the first month (15 December 2018 - 20 January 2019) is conducted using the atmospheric analysis in its 3-dimensional variational (3DVar, \citet{todling2018}) configuration, which is computationally more efficient. The remainder of the spinup simulation (20 January 2019 - 15 March 2019) is conducted using the atmospheric analysis in its Hybrid-4DEnVar configuration. At the transition point, the deterministic assimilation component and the land surface states are initialized with outputs from the 3DVar simulations.  Both ADAS configurations utilize the same version of the GEOS AGCM and assimilate the same set of atmospheric observations.  
 
  \subsection{Validation Data and Metrics}
  \label{ssec:validation}
  
  The evaluation of the impact of SMAP DA on atmospheric forecasts is split into two components: (1) a focused evaluation of the TC analysis and forecast skill using tailored metrics to assess the TC compactness, TC intensity, and TC track and (2) an evaluation of the near-surface atmospheric forecast skill over the entirety of the study domain.
  
 Forecasts of screen-level (2-meter) temperature and relative humidity are evaluated against the corresponding fields from the European Center for Medium-Range Weather Forecasts' (ECMWF) Integrated  Forecast System (IFS) operational weather analysis. The IFS land analysis includes the assimilation of soil moisture estimates from the Advanced Scatterometer (ASCAT) and the Soil Moisture and Ocean Salinity (SMOS) mission \citep{derosnay2013, munoz2019} and -- more importantly --  routinely assimilates station observations of 2-m temperature and humidity \citep{drusch2007}. It is thus considered an appropriate reference to evaluate our forecasts of 2-m temperature and relative humidity. 
 
 For the more TC-focused forecast evaluation, we compute the TC forecast intensity error and the TC forecast track error against the IBTrACS dataset. IBTrACS collects TC data from multiple operational weather forecasting agencies to create a data set of the best observed track and central pressure. In the case of TC Idai, the data used stem from the M\'{e}t\'{e}o France La R\'{e}union subdivision. 
 
 When assessing the TC track error we take into account its two components in the cross-track and along-track direction. Track error is a vector quantity whose two components can take positive or negative values. Positive (negative) cross-track errors indicate that the forecast track is to the right (left) of the observed track. Similarly, for the along-track error positive (negative) values indicate a forecast track that is ahead (behind) of the observed track. Given this and the fact that there are multiple forecasts per lead time, we compute the absolute value of individual forecast errors before averaging all forecasts  for a given lead time and for a given track error component. This approach is taken to avoid compensating errors in the total estimates. 
 
Finally, we assess the compactness of a TC using the wind speed radius, defined as the radius beyond which the wind speed drops below a given threshold value. Typical choices are the 50-knot and 34-knot wind speed radii, R50 and R34.

 \section{Results: SMAP DA Impact on TC Idai Forecasts}
 \label{sec:results_tc}
 
 The focus of our investigation is to identify and quantify to what extent constraining the GEOS land surface initialization with SMAP observations can improve forecasts of TC Idai. That is, we evaluate the model forecasts - CNTRL\_FCST and SMAP\_FCST - initialized from CNTRL\_ANA and SMAP\_ANA, respectively (Table \ref{tab:experiment_overview}) and assess skill differences in forecasts of the near-surface atmosphere (section \ref{sec:results_tc}\ref{ssec:forecasts_2m}) as well as forecasts of the TC Idai intensity and track (section \ref{sec:results_tc}\ref{ssec:forecasts_Idai}). 
 
 \subsection{Forecasts of near surface model states}
 \label{ssec:forecasts_2m}
 
 \begin{figure}[t]
  \noindent\includegraphics[width=\textwidth,angle=0]{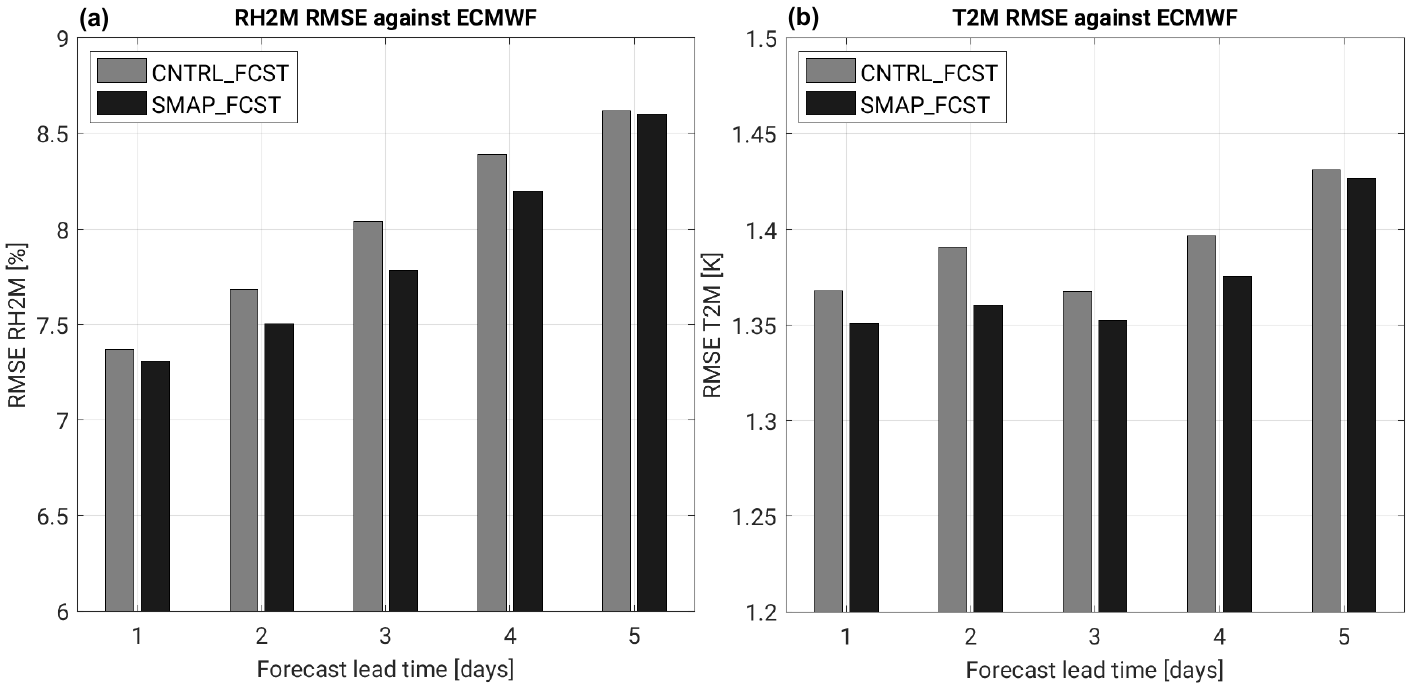}\\
  \caption{RMSE of forecast RH2M (a) and forecast T2m (b) computed against the corresponding fields from the ECMWF IFS operational analysis as a function of forecast lead time for CNTRL\_FCST (gray bars) and SMAP\_FCST (black bars)}.
  \label{figure:rmse_rh2m_t2m_bar}
\end{figure}
 
 First, we assess the model's ability to forecast near surface atmospheric states, specifically the 2-m temperature (T2M) and relative humidity (RH2M), as evaluated against the ECMWF IFS operational analysis (section \ref{sec:data_and_methodology}\ref{ssec:validation}). Figure \ref{figure:rmse_rh2m_t2m_bar} shows the RMSE of RH2M and T2M from the CNTRL\_FCST and SMAP\_FCST evaluated against the ECMWF IFS analysis as a function of the forecast lead time and averaged over the study domain (including land and ocean grid cells). For RH2M (Figure \ref{figure:rmse_rh2m_t2m_bar}(a)),  SMAP\_FCST shows a consistently better skill against the ECMWF analysis than CNTRL\_FCST, with an average difference of 0.13\% and a maximum difference of 0.25\% at a 4-day lead time. The skill difference between CNTRL\_FCST and SMAP\_FCST appears to remain approximately consistent up to and including the 4-day lead time despite increasing total RMSE for both experiments. At a lead time of 5 days the skill of both experiments starts to converge, which is expected as the influence of differences in the forecast initialization weakens over time, consistent with other work \citep[e.g.][]{Prive2013, Cucurull2021, Prive2022}. 
 
 For T2M (Figure \ref{figure:rmse_rh2m_t2m_bar}(b)), SMAP\_FCST also shows a consistently better skill against the ECMWF data than CNTRL\_FCST, with an average difference of 0.02K and a maximum difference of 0.03K at a 2-day lead time. As for RH2M, the skill of the two experiments starts to converge at a lead time of 5 days, again indicating a weakening of the influence of the initialization.
 
 For both RH2M and T2M, the skill differences between CNTRL\_FCST and SMAP\_FCST are largest at lead times beyond 1 day. This might indicate that the longer memory of land surface states comes into effect at longer lead times, as the land becomes a more important source of predictability than the atmosphere. 
 
Overall, the results indicate that the changes to the land surface resulting from the SMAP DA are propagated into the near surface atmosphere via the model physics, where they lead to improved forecasts of atmospheric states. 
 
 \subsection{Analysis and Prediction of TC Idai}
 \label{ssec:forecasts_Idai}
 
 The next and crucial step for this study is to assess how the assimilation of SMAP observations impacts the ability of the GEOS AGCM to represent the structure and predict the evolution of TC Idai. Unlike the evaluation of near-surface atmospheric forecasts, this assessment does not encompass the entire study domain, but is instead directed at TC-centric metrics. Specifically, we focus on three aspects of the TC skill: TC compactness, and forecast TC intensity and TC track.  
 
  \begin{figure}[t]
  \centering
  \noindent\includegraphics[width=0.6\textwidth,angle=0]{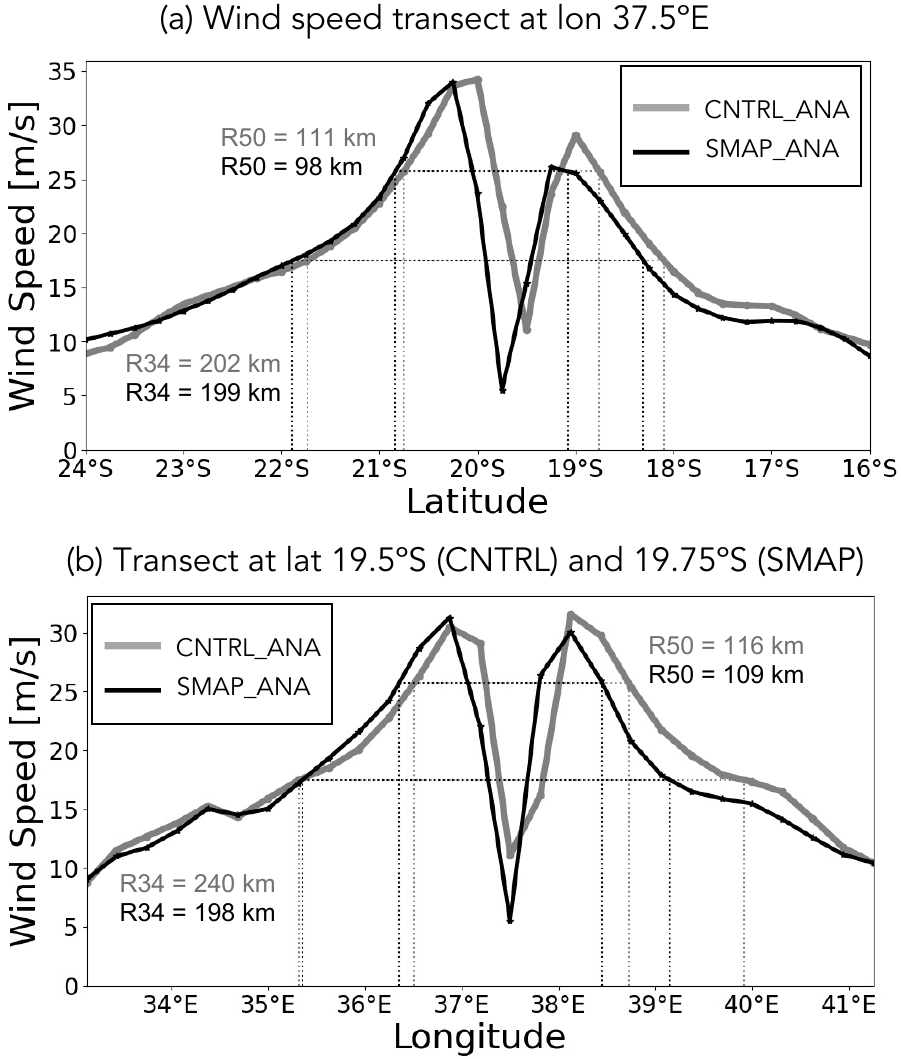}\\
  \caption{850-hPa Wind speed transects across (a) 37.5 degree East longitude and (b) 19.5 degree South latitude for CNTRL\_ANA (gray lines) and 19.75 degree South latitude for SMAP\_ANA (black lines) on 14 March at 00 UTC (TC Idai peak intensity). Also displayed are the corresponding average wind speed radii at 34 knots (R34) and 50 knots (R50).}
  \label{figure:wind_speed_transect}
\end{figure}
 
 As a first step, we focus on evaluating the TC compactness in the analysis. This step serves to establish that the representation of the TC structure is improved in the analysis, which increases confidence in any subsequent assessment of forecast skill improvements. Most operational numerical weather prediction models tend to overestimate the size of TC inner-cores relative to observations, which is a limitation of the spatial resolution at which these models are run. Thus, being able to estimate a more compact TC with a better defined, narrower inner core would be a step in the right direction. Here we assess the TC compactness using the wind speed radius (section \ref{sec:data_and_methodology}\ref{ssec:validation}).
 
 Figure \ref{figure:wind_speed_transect} shows the zonal and meridional wind speed transects for CNTRL\_ANA and SMAP\_ANA at a fixed longitude (37.5 degree East) and latitude (19.5 degree South for CTNRL\_ANA and 19.75 degree South for SMAP\_ANA), respectively, for 14 March at 00 UTC when Idai reached peak intensity. While both representations overestimate the TC size, the SMAP\_ANA is consistently superior.  SMAP\_ANA  reaches a minimum wind speed of $\approx$ 5 m/s at the core, noticeably lower than the wind speed minimum of $\approx$ 10 m/s seen in CNTRL\_ANA, which is a desirable result indicating a more distinctive eye-like feature.  The R50 and R34 metrics show  that in all cases the wind speed radius in SMAP\_ANA is smaller than that in CNTRL\_ANA. For R50, the zonal and meridional wind speed radius differences are 13 km and 7 km, respectively, corresponding to a wind speed radius decrease of 12\% and 6\% as a result of the SMAP assimilation. For R34, the zonal and meridional wind speed radius differences are 3 km and 42 km, respectively, meaning that the assimilation of SMAP leads to a storm that is by 1\% narrower in the zonal direction and 18\% narrower in the meridional direction. The assimilation of SMAP observations thus leads to a more compact TC with a narrower inner-core, which is a highly desirable result. It should be noted that while results here are shown for the time of TC Idai's peak intensity, the wind speed radius in SMAP\_ANA was found to be smaller than that in CNTRL\_ANA at all times when then wind speed radius could be estimated for both experiments (not shown).

  \begin{figure}[t]
  \centering
  \noindent\includegraphics[width=0.6\textwidth,angle=0]{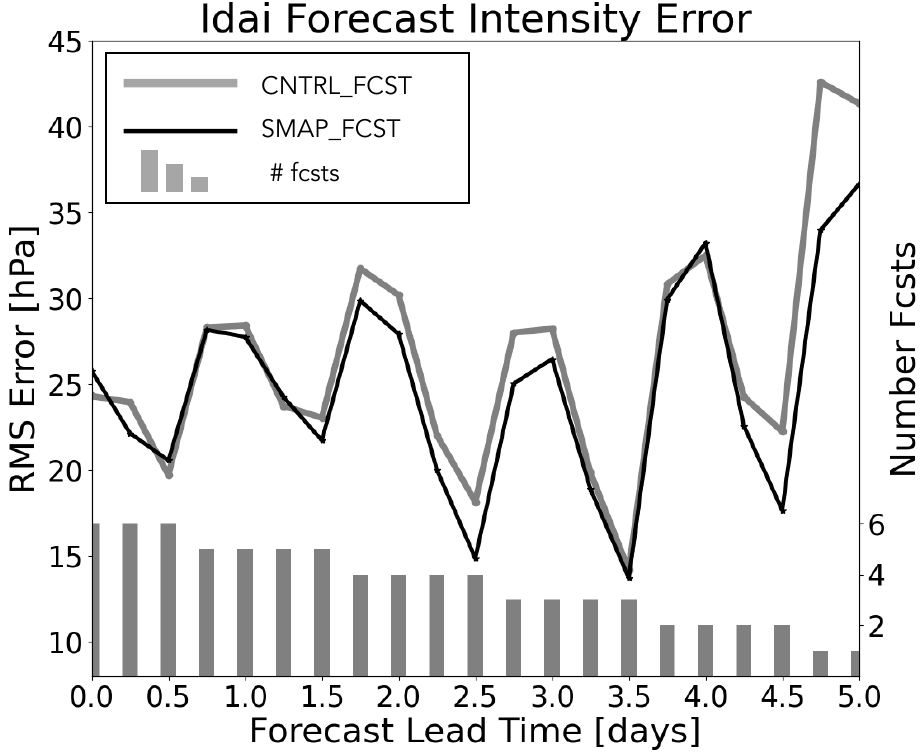}\\
  \caption{Forecast intensity error for TC Idai as a function of the forecast lead time for CNTRL\_FCST (gray line) and SMAP\_FCST (black line). Also shown are the number of forecasts contributing to the error estimate (gray bars), which is identical for CNTRL\_FCST and SMAP\_FCST.}
  \label{figure:TC_Idai_intensity_error}
\end{figure}
 
 Next, we assess the impact of SMAP DA on the forecast intensity error, which is defined as the difference of the forecast central pressure with respect to the IBTrACS best track data (section \ref{sec:data_and_methodology}\ref{ssec:validation}). Figure \ref{figure:TC_Idai_intensity_error} shows the forecast intensity error as a function of lead time for CNTRL\_FCST and SMAP\_FCST. The forecast intensity error in SMAP\_FCST and CNTRL\_FCST is similar for forecasts of 1-day lead time and starts to diverge at 1.5-day lead time, with generally lower errors in SMAP\_FCST. The largest differences occur at forecast lead times of 1.5 days (36 hours) to 3 days (72 hours), when the forecast intensity error in SMAP\_ANA is lower by 1.8 hPa on average and and by 3.5 hPa at the point of maximum difference (2.5-day lead time). This corresponds to an average forecast error reduction of 8\% and a maximum error reduction of 23\% as a result of assimilating SMAP observations. Large differences can also be seen beyond 4-day lead times, but the confidence in intensity error estimates at such large lead times is reduced, because of the low number of contributing forecasts compared to lower lead times. Overall, the results clearly indicate an improvement in the intensity forecast skill as a result of SMAP DA.

\begin{figure}[t]
  \centering
  \noindent\includegraphics[width=\textwidth,angle=0]{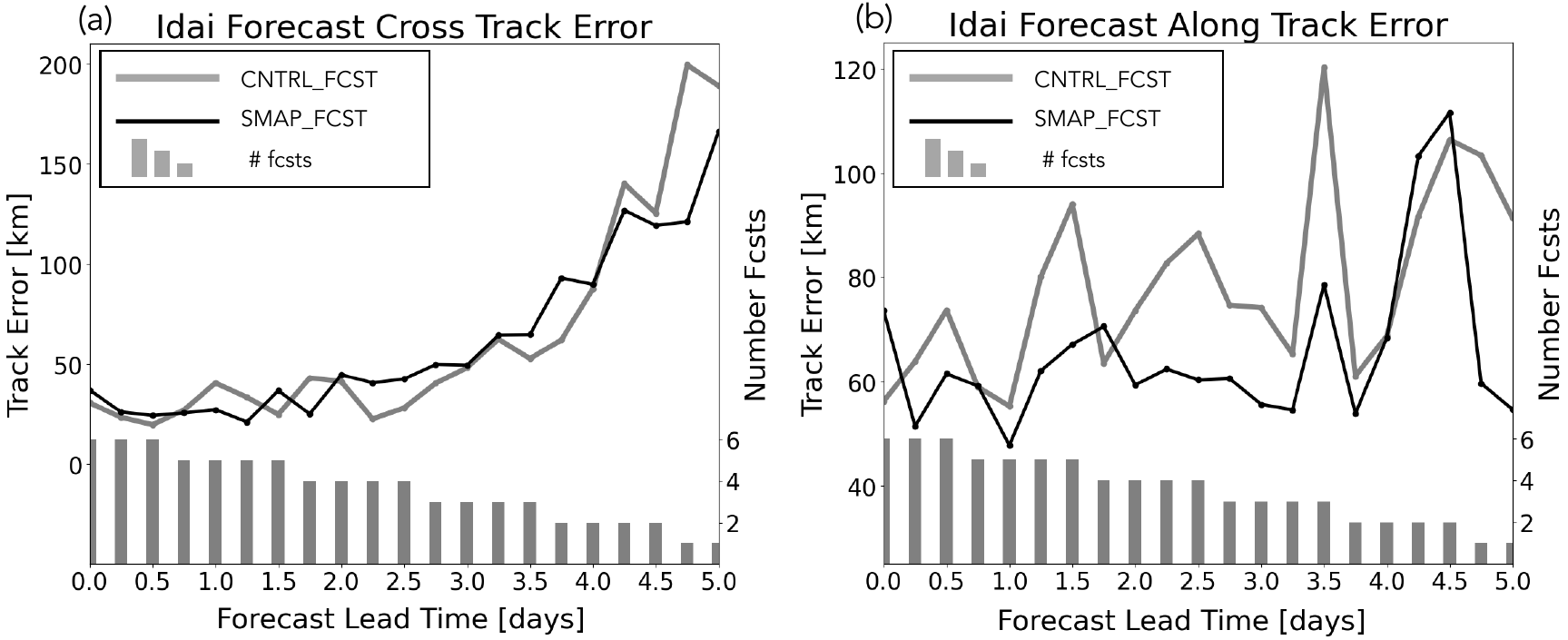}\\
  \caption{TC Idai forecast mean absolute track errors as a function of lead time for CNTRL\_FCST (gray lines) and SMAP\_FCST (black lines) for the (a) cross-track errors and (b) along-track errors compared to the observed track. Also shown are the number of forecasts contributing to the error estimates (gray bars). }
  \label{figure:TC_Idai_track_error}
\end{figure}

 The final aspect of assessing the TC forecast skill is the TC track error, split into its cross-track and along-track components. As for the intensity, the TC track error is evaluated against the IBTrACS best track estimate (section \ref{sec:data_and_methodology}\ref{ssec:validation}). Figure \ref{figure:TC_Idai_track_error} shows the mean absolute cross-track and along-track error of the TC Idai forecasts from CNTRL\_FCST and SMAP\_FCST as a function of the forecast lead time. 
 
 As expected, the mean absolute cross-track errors in CNTRL\_FCST and SMAP\_FCST increase with forecast lead time (Figure \ref{figure:TC_Idai_track_error}a). Cross-track error differences between both experiments are small, leading to the conclusion that the impact of SMAP DA on the forecast cross-track error is neutral. An examination of individual errors (not shown) revealed that in both experiments most errors are negative, indicating that the forecasts from both experiments are biased to the left of the observed track.

 The mean absolute along-track errors in SMAP\_FCST are generally smaller than those in CNTRL\_FCST (Figure \ref{figure:TC_Idai_track_error}b), indicating that the SMAP assimilation reduces the along-track error and moves the forecast track closer to the observed track. The along-track error is 12 km smaller for SMAP\_FCST than for CNTRL\_FCST over the entire period, with the largest error reduction of 41 km occurring at 3.5-day lead time. This corresponds to an average error reduction of 15\% and a maximum error reduction of 34\% relative to the CNTRL\_FCST along-track error. The corrective effect of the SMAP observations is clearly present up to a forecast lead time of 3.5 days, beyond which there are too few forecasts contributing to the metric to draw meaningful conclusions. While the along-track error differences between CNTRL\_FCST and SMAP\_FCST are based on relatively few forecasts, they are consistent, indicating that there is a real benefit from the assimilation of SMAP observations. An investigation of individual forecast errors (not shown) revealed that most forecasts in both experiments  have a positive forecast along track error, indicating that both experiments predict a track that is ahead of the observed track, with SMAP DA allowing a systematic correction of this error. 
 
 From the above results it can be concluded that -- as hypothesized -- the assimilation of SMAP Tbs into a numerical weather prediction model significantly benefits the representation and forecast skill for TC Idai. These improvements are seen in key aspects of TC estimation, including TC compactness, and predicted TC intensity and TC track. The improvements of SMAP DA on the TC Idai prediction skill are largest in the 36 to 72 hour time frame, suggesting that the benefit from SMAP is largest when  the initial predictive skill from atmospheric states is reduced and the land, with its longer memory, becomes increasingly important as a source of predictability. 
 
 \section{Results: Underlying Surface Mechanisms}
 \label{sec:results_surface_mechanisms}
 
 So far our investigation has focused on quantifying the impact of SMAP DA on the model's ability to simulate TC Idai when it is near or over land. As a next step, we shift our focus to identifying the mechanisms by which the forecast skill improvements seen in section \ref{sec:results_tc} are achieved. Unlike section \ref{sec:results_tc}, the following investigations make use of the model analyses (CNTRL\_ANA and SMAP\_ANA) to examine how the assimilation of SMAP impacts the surface states and fluxes and how these changes are propagated into the atmosphere. 
 
 \subsection{Surface Analysis}
 
  \begin{figure}[t]
  \noindent\includegraphics[width=\textwidth,angle=0]{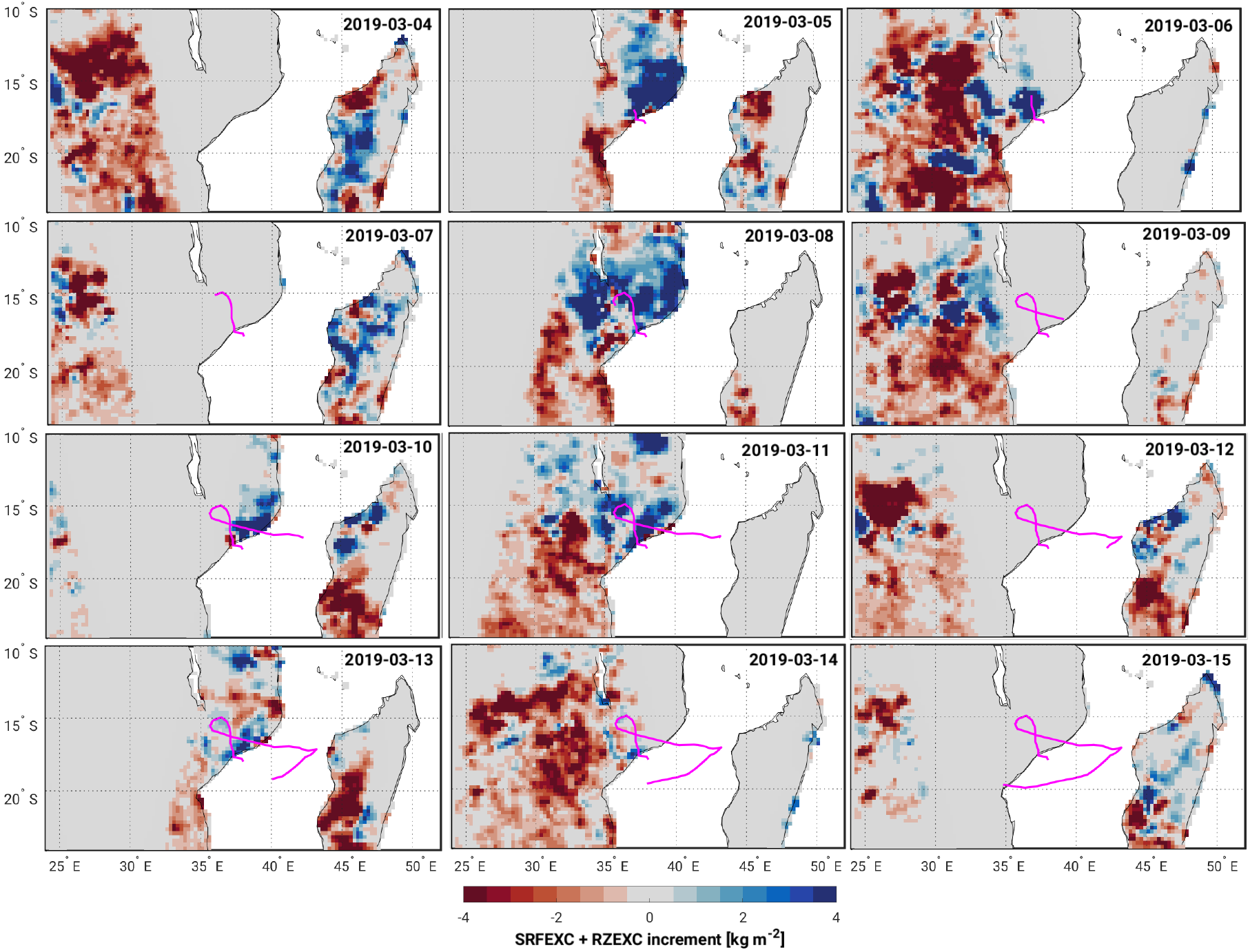}\\
  \caption{Daily average total soil water (SRFEXC + RZEXC) increments (kg m$-2$) throughout the TC Idai life cycle over the TC Idai study domain. Red (blue) colors indicate SMAP DA results in drier (wetter) soil moisture conditions relative to the control run. Gray colors indicate zero (neutral) increments.}
  \label{figure:srfexc_rzexc_incr}
\end{figure}

 We first investigate how the assimilation of SMAP affects the surface states in the land analysis. Figure \ref{figure:srfexc_rzexc_incr} shows the daily average total increments (SRFEXC + RZEXC) applied over the study domain during the TC Idai life cycle. The impact of SMAP is characterized by strong drying increments over Central Mozambique and Eastern Zimbabwe ($\approx$ 13\textdegree S - 25\textdegree S and 24\textdegree E - 35\textdegree E) that are present for most of the TC's life cycle and only slightly weaken at the end of TC Idai's life (15 March) when the storm and its associated precipitation make landfall in Central Mozambique. Also prominent in Figure \ref{figure:srfexc_rzexc_incr} are wetting increments over Northern Mozambique ($\approx$ 10\textdegree S - 18\textdegree S and 35\textdegree E - 40\textdegree E) and Central Madagascar. These wetting increments are present at the beginning of the TC Idai life cycle (4 and 5 March) and increase in strength over Northern Mozambique as the storm moves and precipitates over the region (6 - 9 March). An investigation of the individual SRFEXC and RZEXC plots (not shown) reveals that the strong drying from SMAP DA appears to be predominantly a surface feature, whereas the wetting increments in Northern Mozambique and Central Madagascar are relatively more pronounced in RZEXC.

\begin{figure}[t]
  \noindent\includegraphics[width=0.9\textwidth,angle=0]{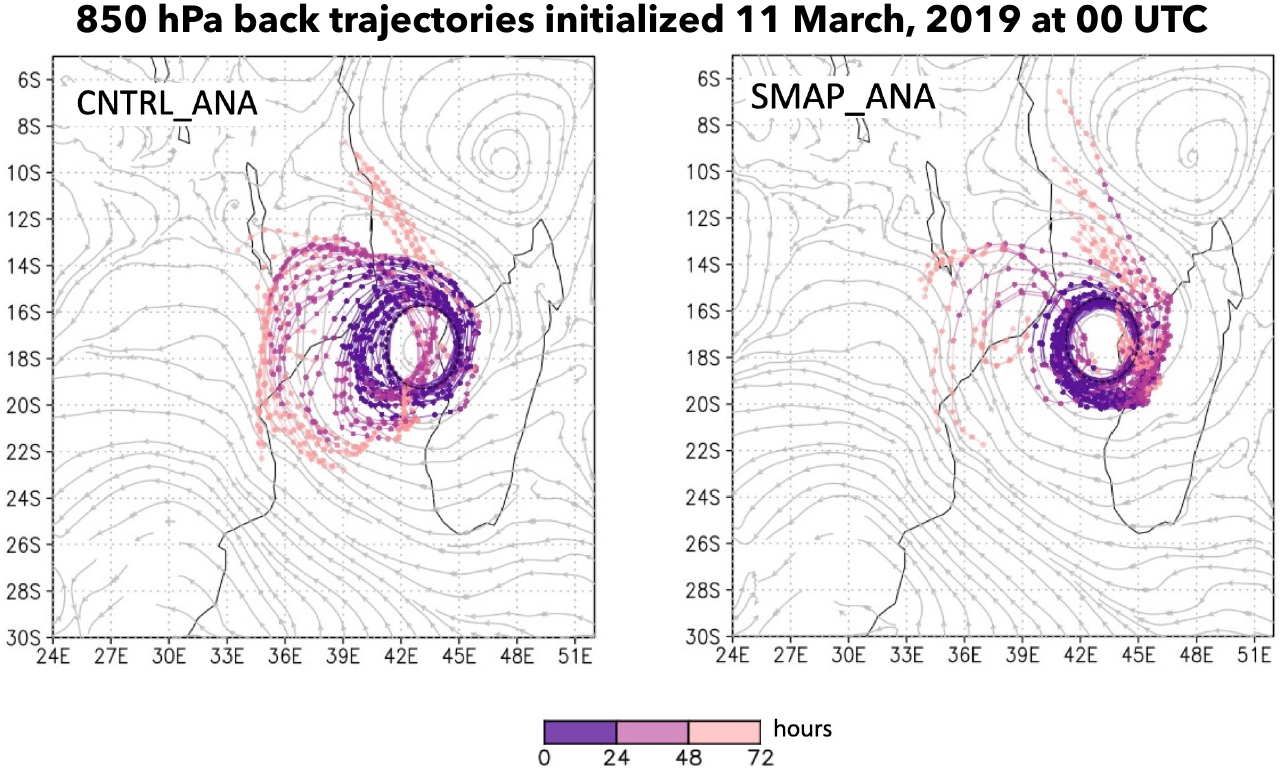}\\
  \caption{Back trajectories initialized on 11 March at 00 UTC at 850 hPa for CNTRL\_ANA (left panel) and SMAP\_ANA (right panel) color coded to their time history in intervals of 24 hours. Gray contours represent the underlying 850 hPa wind fields in each simulation.}
  \label{figure:back_trajectories}
\end{figure}

All of the soil water increments  seen in Figure \ref{figure:srfexc_rzexc_incr} and their resulting impact on soil moisture (not shown) can influence the broader atmospheric circulation that then impacts the evolution of TC Idai. However, the focus of this study is to quantify the impact of land areas that directly influence TC Idai throughout its life cycle. We therefore isolated the land areas in the vicinity of the storm that influence the storm's circulation using the back trajectory analysis algorithm provided through the George Mason University GrADS script library \citep{gmu}. For each day at 00 UTC during the TC life cycle, we define a circle with a 200-km radius around the center of the storm to delineate the storm's outer core, separately for each experiment (CNTRL\_ANA and SMAP\_ANA). Next, 36  points are placed on this circle at 10\textdegree -spacing and used as starting points for the back trajectory calculations. Each back trajectory is traced back in time for 72 hours using the 850 hPa simulated wind fields. A sample of these back trajectories for 11 March 00 UTC is shown in Figure~\ref{figure:back_trajectories}. Grid cells where the back trajectories intersect land are then used to construct a land mask -- separately for each day and experiment -- that isolates the land areas that directly influence the evolution of TC Idai on a given day.  
 
   \begin{figure}[t]
   \centering
  \noindent\includegraphics[width=0.9\textwidth,angle=0]{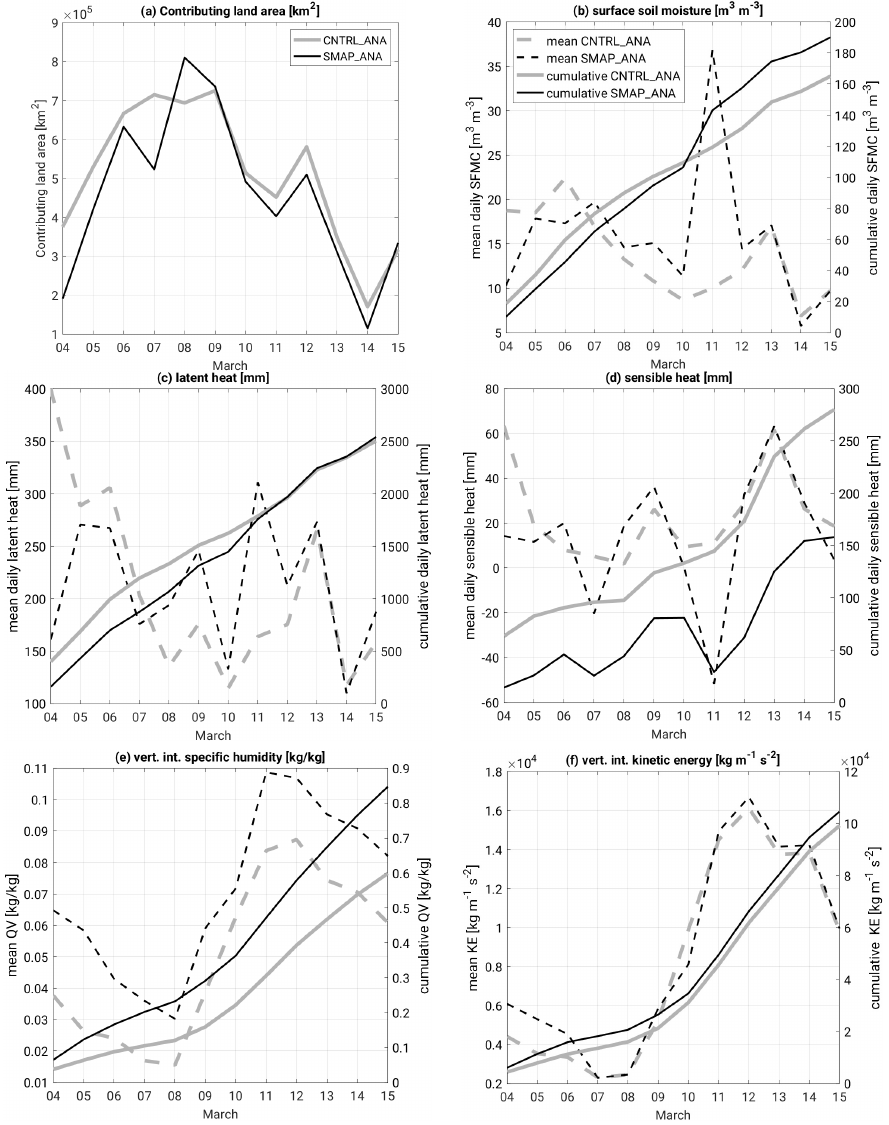}\\
  \caption{Mean (dashed lines; left axes) and cumulative sum (solid lines; right axes) of (a) contributing land area ($km^2$), (b) surface soil moisture ($m^3 m^{-3}$), (c) latent heat flux ($mm$), (d) sensible heat flux ($mm$) (e) specific humidity ($kg/kg$) and (f) kinetic energy ($kg m^{-1} s^{-2}$)
  during TC Idai's life cycle (March 4 - 15) for (black) SMAP\_ANA and (gray) CNTRL\_ANA.}
  \label{figure:mean_cumul_all}
\end{figure}

 Next, we use the land mask obtained from the back trajectories to examine the temporal evolution of the land surface states and fluxes over the land areas that directly impact TC Idai. To this end we compute the spatially averaged states and fluxes for each time step using the state/flux value from the time when a back trajectory intersects a given land grid cell. Land grid cells that are intersected by multiple back trajectories are assigned a higher weight than those intersected by a single trajectory only. 
 
 The total land area that contributes to each average is shown in Figure~\ref{figure:mean_cumul_all}a, and the average and cumulative sum of the area-averaged surface fluxes and states during TC Idai's life cycle are shown in Figure~\ref{figure:mean_cumul_all}b-d. The differences in the average surface states and fluxes between SMAP\_ANA and CNTRL\_ANA vary on a day-to-day basis and the temporal evolution of these differences does not appear to correspond to the temporal evolution of the total land area contributing to the area-average (Figure~\ref{figure:mean_cumul_all}a), suggesting that the state and flux difference instead reflect differences in the land surface conditions. 
 The average surface soil moisture (Figure \ref{figure:mean_cumul_all}b) shows higher soil moisture values in CNTRL\_ANA during the first days of the storm (4 - 6 Mar), whereas the soil moisture is higher in SMAP\_ANA during 7 - 12 Mar. During the last days of the TC Idai life cycle (13 - 15 Mar) both experiments show very similar behavior. The differences in the daily average soil moisture are also reflected in the cumulative behavior, which shows higher cumulative surface moisture in CNTRL\_ANA for 4 - 10 Mar, which then switches to a higher cumulative soil moisture in SMAP\_ANA from 11 Mar onward. The switch in the cumulative differences appears to be largely driven by a much wetter soil moisture in SMAP\_ANA on 11 Mar. Despite both experiments seeing a similar total land area on that day (Figure \ref{figure:mean_cumul_all} (a)), an investigation of the back trajectories (Figure \ref{figure:back_trajectories}) reveals that SMAP\_ANA is predominantly influenced by the wet areas of northern Madagascar (Figure \ref{figure:srfexc_rzexc_incr}), whereas CNTRL\_ANA is impacted by the land areas of central and northern Mozambique, which show a mix of wet and dry differences. Notably, 11 Mar is the day Idai experienced rapid intensification (section \ref{sec:tc_idai}), hinting that the wet surface conditions over Madagascar may have played a role in this.

  Both the average and cumulative latent heat flux (Figure \ref{figure:mean_cumul_all}c) mirror the qualitative behavior observed for soil moisture, reflecting the connection between soil moisture and latent heat flux. However, the relative differences in the fluxes between both experiments diverge from those observed for soil moisture, as the latent heat flux differences are additionally driven by radiative forcing differences owing to departures in cloud cover in both experiments. The differences in average sensible heat flux (Figure \ref{figure:mean_cumul_all}d) are less coherent, but do show a dip in the sensible heat flux in SMAP\_ANA on 11 Mar, which correspond well to a wetter land surface that evaporates more and is thus cooler. A strong signal of consistently higher sensible heat flux in CNTRL\_ANA compared to SMAP\_ANA is apparent in the cumulative fluxes. This difference is present on the day of TC genesis (4 Mar) and -- owing to the mostly small daily average differences between both experiments -- is persisted throughout the TC life cycle.

  \subsection{Atmospheric Analysis}
  
  Next, we examine to what extent SMAP DA influences the simulation of TC Idai. Here we are focused on the analyses, SMAP\_ANA and CNTRL\_ANA. Aiming to assess differences in the entirety of the storm, this analysis is focused on two metrics: (1) vertically integrated specific humidity (QV), used here as a proxy for the moisture content of the storm, and (2) vertically integrated kinetic energy (KV), used here as a proxy for the energy contained in the storm.  The latter is computed by vertically integrating \(\frac{1}{2}*\rho*v^{2}\), where $\rho$ is the air density and $v$ is the wind speed. Both metrics are computed as daily quantities that are spatially averaged within the area surrounding the storms center using a TC wind speed threshold of 17 m s\textsuperscript{-1}.
  
  Figure \ref{figure:mean_cumul_all}e shows the spatially averaged QV throughout the TC life cycle for SMAP\_ANA and CNTRL\_ANA. Both experiments show the same qualitative behavior for the daily mean QV with an initial high moisture content that gradually decreases during 4-7 March as the storm loops over Northern Mozambique. This moisture loss is typical for a TC over land. In both analyses, QV increases again starting 9 March, after Idai re-emerges in the Mozambique channel and is provided with moisture from the ocean. Quantitatively, the storm's moisture content is consistently  higher in SMAP\_ANA compared to CNTRL\_ANA. Generally, we expect QV to be dominated by the moisture uptake from the ocean and thus the fraction of the TC circulation that is over open water. Nevertheless, we see clear evidence of the land influence and thus the impact of SMAP DA in Figure \ref{figure:mean_cumul_all}e. In particular on 11 Mar, when the land areas with direct influence on the storm show a significantly increased soil moisture and latent heat flux in SMAP\_ANA compared to CNTRL\_ANA, we see a clear increase in the QV difference between SMAP\_ANA and CNTRL\_ANA, reflecting the moisture injection from the land in SMAP\_ANA. We also note that the QV difference between both experiments is present at the beginning of Idai's life cycle on 4 March, suggesting that at least some of the difference observed in Figure \ref{figure:mean_cumul_all}e result from the cumulative influence of SMAP DA during the 2.5 months prior to TC Idai, which can alter the antecedent soil moisture conditions and lead to differences in the atmospheric conditions prior to TC genesis.
  
  For the KV, the daily mean difference between SMAP\_ANA and CNTRL\_ANA is less pronounced (Figure \ref{figure:mean_cumul_all}f). Both experiments again show a similar qualitative behavior that mirrors the behavior observed for QV. That is, at first KV slowly decreases as the storm loops over land and its energy dissipates. Once TC Idai re-emerges over the Mozambique channel on 9 March, KV increases reaching a peak value around 12 March. Cumulatively, the total energy in SMAP\_ANA is higher compared to CNTRL\_ANA, although generally the relative differences are smaller than those seen for QV. The KV differences between both experiments do not have a strong correspondence to the sensible heat flux or the land area under the storm's circulation, suggesting that factors other than the land are the main drivers of KV.  
  
  Both the higher moisture content and higher total energy of TC Idai as simulated in SMAP\_ANA compared to CNTRL\_ANA align better with the known devastating nature of TC Idai.

 \section{Summary and Conclusions}
 \label{sec:conclusions}
 
Our study investigated the impact of assimilating SMAP observations on the analysis and forecast skill of TC Idai using a global numerical weather prediction model. To this end we constructed an OSE framework that systematically assessed the impact of SMAP DA on the analysis and prediction of TC Idai, which affected Mozambique and Madagascar in March 2019.

Considering the entirety of our study domain, we observe that SMAP DA tends to dry the soil moisture relative to a control run in Central Mozambique, while wetter conditions are observed over Northern Mozambique and Northern Madagascar. However, when isolating the land areas that directly influence TC Idai and taking into account the circulation patterns over these areas, we see that SMAP DA leads to wetter soil moisture conditions over the land areas that are relevant to the storm for the majority of the TC life cycle. The elevated soil moisture over the contributing land areas is also reflected in an increased latent heat flux in response to SMAP DA, which indicates an increased moisture flux to the atmosphere. 

Accordingly, our investigation of column-integrated atmospheric states shows that the assimilation of SMAP leads to a TC analyzed representation that has  higher total moisture content and total energy compared to the analysis in the control run without SMAP DA. We also find that the analyzed TC compactness, as measured by the wind speed radius, is improved in the analysis with SMAP DA relative to the control run. Global models at 1/4-degree resolution such as the one used here unavoidably under-estimate TC intensity and the analyzed TCs tend to be less compact than indicated by observations. Therefore, the SMAP assimilation forces the model in the right direction, leading to a storm that is more energetic and more compact. 
 
The improvements from SMAP DA in the analyzed TC representation are also propagated into forecasts of TC Idai, with a beneficial impact seen for the forecast TC intensity and TC track. 
Effective forecast error reductions are obtained for the predicted TC Idai intensity (as measured by minimum sea level pressure). Evaluated against observed TC intensity, the intensity error of forecasts initialized from an analysis with SMAP DA is reduced by 8\% on average with a maximum error reduction of 23\% compared to that of forecasts initialized without SMAP assimilation. 
Finally, SMAP DA also shows a beneficial impact on forecasts of the TC Idai track when evaluated against the observed best track. While the impact on the cross-track error component is neutral, the along-track error is reduced by 15\% on average and 34\% at the maximum, in forecasts initialized from an analysis with SMAP DA compared to a control run without SMAP DA, indicating a more accurate propagation speed, consistent with the fact that TC speed over land is strongly affected by surface processes.

In terms of forecast skill over the entire study domain, the impact of SMAP DA is generally positive.  Changes to the land surface resulting from the SMAP DA are propagated into the near surface atmosphere via the model physics, where they lead to improved forecasts of near surface atmospheric states when evaluated against the observation-constrained ECMWF operational analysis. The maximum error reduction in the screen-level temperature and humidity is on the order of 0.25\%, which is small but not negligible.

Across all the TC forecast skill metrics used here, the improvements from SMAP DA are largest in the 36 to 72 hour time frame. This suggests that the predictability of forecasts at lower lead times may be dominated by short-term convective processes which cannot be fully resolved by this model, while the land -- with its longer memory -- gains in importance as a source of predictability at longer lead times. 

Despite these very positive results, more work is needed to better understand the full extent to which SMAP DA is able to improve TC predictions. Here we have identified a very direct impact of SMAP DA, whereby an increased surface soil moisture leads to an increased latent heat flux (which connects the land surface to the atmosphere), resulting in an increased total atmospheric moisture and energy content in TC Idai. However, we have also noted that differences between our SMAP DA and control experiments exist at the beginning of the study period and are persisted throughout the TC life cycle. This suggests that the impact of SMAP DA on the TC prediction may not be confined to the period of the storm and that some of the impact may stem from the 2.5-month spin-up period during which SMAP observations were assimilated prior to the storm.  However, an investigation of this potentially broader effect of SMAP DA on TCs is reserved for a future study. A larger set of TC simulations is likely required to address this question. 

Overall, the results highlight that -- as hypothesized -- the assimilation of SMAP observations into a global numerical weather prediction model can provide a meaningful improvement of TC predictions. This is a crucial step towards our ability to mitigate the impact of these disastrous weather events. It also highlights an important aspect of evaluating the impact of land DA in the context of numerical weather prediction. While global skill assessments are necessary, they can mask the true value of land DA in operational systems (see section \ref{sec:introduction}). Instead, a more event-oriented approach may help to better characterize the value of land DA in cases when the land has an impact on the atmosphere, especially for extreme events when accurate predictions are critical for safeguarding human lives.

\clearpage
\acknowledgments
The authors gratefully acknowledge support by Dr. Jared Entin (NASA HQ) through NASA grant 80NSSC21K0323. Computational resources were provided by the NASA High-End Computing program through the NASA Center for Climate Simulation.

%
%
\datastatement
Documentation, tools, and methods used in this study are available through the NASA GMAO Github page (\url{https://github.com/GEOS-ESM}) or are otherwise available through the resources cited in this manuscript. The SMAP data used here (DOI: 10.5067/JJ5FL7FRLKJI) are publicly available through the National Snow and Ice Data Center (\url{https://nsidc.org/data/smap}). The dataset on which this paper is based is too large to be retained or publicly archived with available resources, but can be readily reproduced with the publicly available tools and input data mentioned above. Documentation and methods used to support this study are available from the lead author at the NASA Global Modelling and Assimilation Office.

%






%



 \bibliographystyle{ametsocV6}
 \bibliography{references}

\end{document}